# On-chip frequency tuning of fast resonant MEMS scanner

Paul Janin[1], Deepak Uttamchandani[1], and Ralf Bauer[1,*]

[1] Department of Electronic and Electrical Engineering, University of Strathclyde, Glasgow, UK
[*] Ralf.Bauer@strath.ac.uk

## Abstract

The development and characterisation of a piezoelectric actuated high-frequency MEMS scanning mirror with on-chip frequency tuning capability is reported. The resonant scanner operates at frequencies in excess of 140 kHz, generating scan angles of 10° and 6° for two orthogonal movement modes with 40 V actuation. On-chip frequency tuning is achieved through electrothermal actuators fabricated adjacent to the mirror main suspension. The electrothermal actuators produce a global and local temperature increase which changes the suspension stiffness and therefore the resonant frequency. A resonance frequency tuning range of up to 5.5 kHz is achieved, with tuning dominant on only one of the two orthogonal scan movement modes. This opens the possibility for precise tuning of a 2D Lissajous scan pattern using a single resonant MEMS scanner with dual orthogonal resonant modes producing full frame update rates up to 20 kHz while retaining the full angular range of both resonant movement modes.

## Introduction

Microelectromechanical System (MEMS) scanning micromirrors have in the last decades made the move from research demonstrations into products such as pico-projectors [1],[2], display applications [3], LIDAR [4],[5], biomedical imaging [6],[7] and smart headlights [8]. Over the years a wide variety of scanner designs with various actuation mechanisms including electrostatic, electrothermal, electromagnetic and piezoelectric actuation have been demonstrated [9]. Each actuation approach has its distinct advantages and disadvantages, but all, in general, yield scan angles ranging from 5° to more than 40° with quasi-static angular positioning or resonant motion in the few kHz to tens kHz range. While especially for projection purposes raster scanning approaches are most commonly used, leveraging the combination of one quasi-static and one orthogonal resonant movement axis, dual-resonant scan approaches have also been demonstrated based on Lissajous scan patterns [10], [11], [12]. The latter allow the use of faster individual axes with potential limitations of control over the full field addressing of a 2D scene. Lissajous patterns can be created in non-repeating or repeating fashion depending on the frequency ratio of the two scan axes, with repeating patterns creating an easier synchronisation with light sources or sensors. While raster scanning approaches are in general limited to full field update rates of hundreds Hz, Lissajous approaches use high scan frequencies in both axes and can allow faster repetition rates of individual points or full field of views. For projection applications this can be achieved by synchronising an excitation source with the instantaneous MEMS position. However, if precise control of update frequency and positioning of scan traces and points is required then control of the frequency ratio of the two scan axes is essential. To gain high density and high frequency frame rates a frequency selection rule has recently been reported [13], [14], with frame rates still strongly dependent on the physical dimensions of the micromirrors due to, in general, high Q factor resonance movements for fast scanning micromirrors. Manufacturing tolerances of specifically high frequency resonance MEMS mean that even 0.1% deviations in the resonance frequency can lead to significant shifts in the desired update rate, and would require adjustment mechanisms for frequency tuning. Approaches for frequency tuning of resonant MEMS micromirrors have so far mostly focused on the removal or addition of material, for example by focused ion beam milling and removal of material from the edges of a scanner mirror plate [15], which has achieved a scan frequency adjustment of 3.5%. While this passive tuning method does not increase the power consumption of the scanner it also does not allow active adjustment during operation, is not reversible and not mass-fabrication compatible.

Additional approaches for active tuning through specific designs on the micromirror silicon chips have been shown for low frequency 1D scanners by using clamping beams on the torsional spring connectors of micromirrors, either through reversible movement of actuation beams [16] or by using local thermal stress effects [17].

In this paper, a piezoelectric scanning MEMS micromirror with tip and tilt frequencies well above 100 kHz and an on-chip frequency tuning mechanism is presented. The mirror, fabricated with a multi-user process, allows tuning of the orthogonal angular movement frequencies by up to 5.5 kHz and 0.55 kHz, or 4% and 0.3% of the respective resonance frequency. The tuning is achieved through on-chip electrothermal actuators with tuning tips which are in close proximity to the scanner main torsion beams. It is shown that while all resonance mode frequencies are tuned at the same time using the electrothermal tuning, an order of magnitude difference in the rate of change allows for adjustment of the ratio between orthogonal movement mode frequencies. The paper reports results on the frequency spread of the scanner design based on measurements from six devices produced in a single fabrication run, followed by evaluation of two designs of electrothermal tuning elements and their individual and combined maximum tuning range characterisation.

## MEMS design

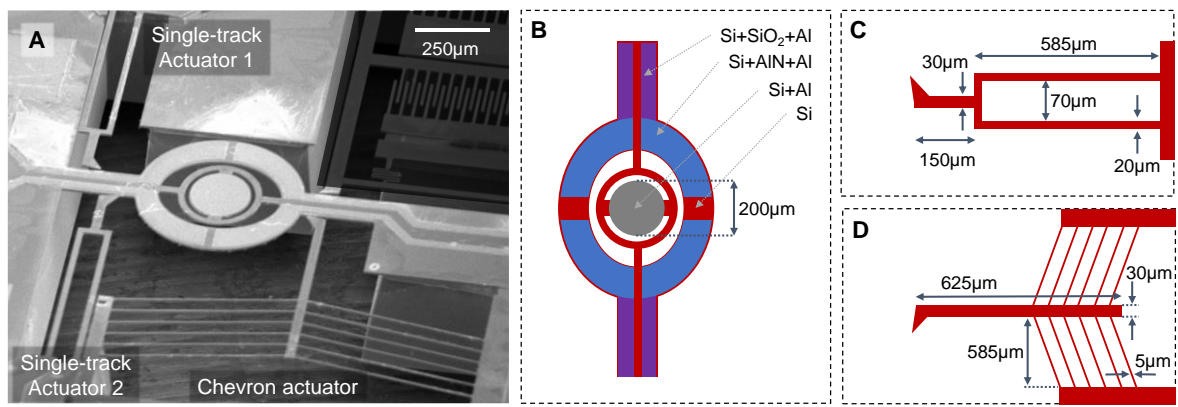

*Figure 1: (A) SEM the fabricated MEMS frequency tuning scanner. (B) Design of the piezoelectric double frame scanner, (C) design of the single-track tuning actuator, and (D) design of the chevron tuning actuator*

The piezoelectric resonant scanning micromirror devices presented here (see Fig. 1) are a variation of MEMS scanners already described in [18], [3], modified to accommodate additional components for frequency tuning. The scanners consist of a 200 µm circular SOI mirror plate coated with 500 nm of aluminium and connected to a gimbal frame with 20 µm width for mirror dynamic deformation reduction. The inner frame is suspended in the middle of an elliptic gimbal by 25 µm wide and 65 µm long beams along the major axis of the ellipse. The elliptic gimbal contains the piezoelectric actuators and itself is suspended and attached to the SOI die by two 100 µm wide and 240 µm long suspension beams. Four independent aluminium nitride (AlN) piezoelectric actuators are placed on the quadrants of the gimbal and electrically connected to the die bonding pads. The actuators are formed with a 500 nm thin film of AlN onto which a 1 µm layer of aluminium is deposited. The aluminium layer serves as an electrode to actuate the piezoelectric layer, with the doped single crystal silicon device layer used as common ground. The scanners are fabricated using a cost-effective multi-user process (PiezoMUMPs offered by Memscap Inc,) which limits the silicon device layer thickness to 10 µm thick silicon and allows process steps for a fully backside released structure.

The frequency tuning is realised through electrothermal SOI actuators with a tip fabricated adjacent to the scanner main suspension beams. The electrothermal actuation produces a global and localized increase in temperature, resulting in a change in suspension beam stiffness and consequently a shift in resonance frequency. Three electrothermal tuning actuators are fabricated, with two using a single-track tuning fork actuator (see Fig. 1C), while the third uses an array of thin beams in a chevron shape (see Fig. 1D) to produce in-plane displacements resulting from thermal expansion and beam buckling. The single-track actuator has an overall length of 735 µm, with a 585 µm long and 20 µm wide current track connecting to a tip of 150 µm length which without actuation is located 5.4 µm from the scanner main suspension beam. Of the two single-track actuators one is placed 110 µm from the scanner substrate connection (identified as "single-track 1" actuator and placed on the opposite side of the main scanner rotation axis as the chevron actuator) while the other actuator is placed 190 µm from it (identified as "single-track 2" actuator and placed on the same side of the main scanner rotation axis as the chevron actuator). The chevron actuator has an overall length of 700 µm, with a central 30 µm wide actuator beam connected to the substrate via six 5 µm wide and 585 µm long beams placed at an 8° rake angle. The tip of the actuator is again located 5.4 µm from the scanner main suspension beam without thermal actuation and 220 µm from the scanner substrate connection.

## Scanner characterisation

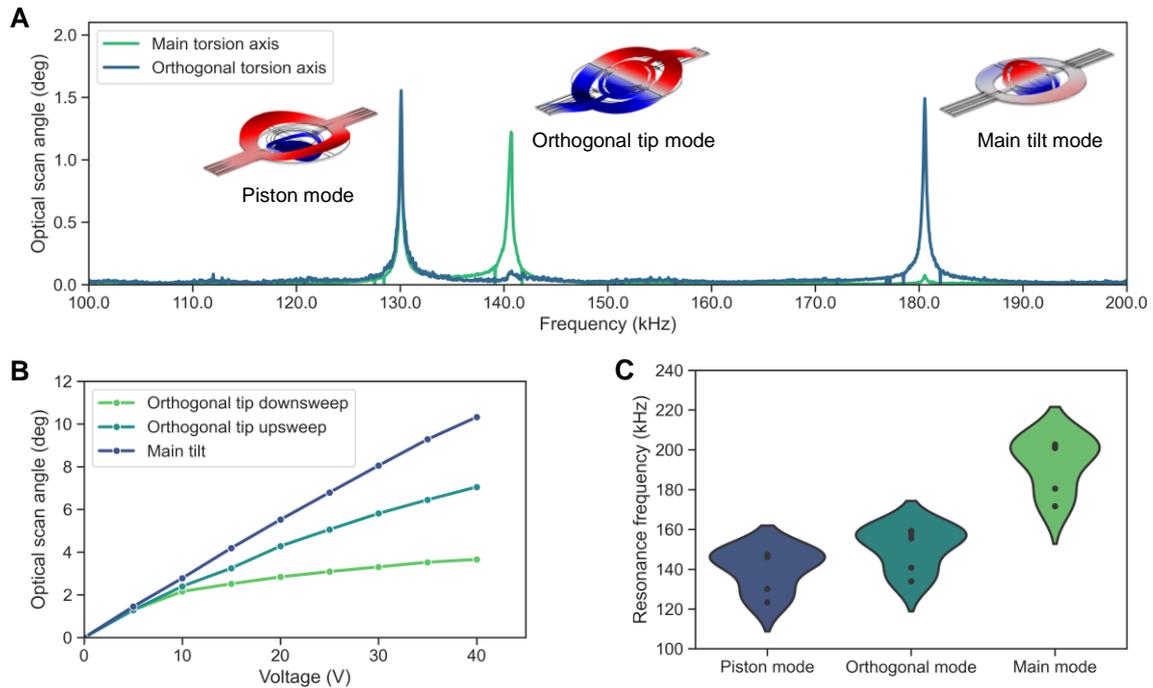

*Figure 2: Frequency response movement behaviour of the scanner without frequency tuning. (A) Frequency response to a 5 Vpp actuation of one of four piezoelectric actuators, measured with a microscope coupled vibrometer with single spot measurement points placed on the circumference of the 200 µm diameter circular mirror plate along the axis of the main torsion beam (green) and orthogonal to the axis of the main torsion beam (blue). Insets show the movement mode shape for each resonance peak. (B) Resonant scan angles for the tip and tilt movements with varying actuation voltage using a single piezoelectric actuator. (C) Distribution of the resonance eigenfrequencies for six mirrors fabricated on the same MEMS multi-user processing run.*

The frequency response of the scanning micromirror is measured using a Polytech OVF 512 Doppler vibrometer coupled to a microscope body. The single point vibrometer response is collected at two positions on the mirror surface, a point at the edge of the mirror in-line with the main suspension beam axis and a point at the edge of the mirror in line with the orthogonal mirror-frame connection. One out of the four piezoelectric actuator is used for the test, with an offset sine-wave signal of 5 $V_{pp}$ and 2.5 V offset applied to the actuator. A frequency sweep with 50 Hz step size is conducted from 100 kHz to 200 kHz, covering the main tip, tilt and piston movement modes (see Fig 2A for the measured angular response together with simulated modeshapes). Three distinct movement modes are visible around 130 kHz, 140 kHz and 180 kHz, with a response on both measurement points for 130 kHz indicating a piston resonance mode, a response on only the main axis at 140 kHz indicating a tip resonance mode, where the axis of rotation is perpendicular to the main suspension beams, and a response on only the transverse axis at 180 kHz indicating a tilt resonance mode, which has the axis of rotation parallel to the main suspension beams of the scanner. The angular response of the two highlighted scanning movement modes for offset sinusoidal actuation voltages up to 40 $V_{pp}$ applied to a single piezoelectric actuator is shown in Fig. 2B. The maximum voltage is conservatively capped at this value for all actuation cases to avoid any potential depolarisation or damage of the piezoelectric actuators. The main tilt movement mode reaches an optical scan angle of 10.3° with a single piezo actuator at 40 $V_{pp}$, while the orthogonal tip movement mode reaches an optical scan angle of 7° with a single piezo actuator when reaching the resonance frequency from a lower frequency point. The tip movement shows considerable hysteresis at this point and when reaching the peak resonance from a higher frequency the maximum optical scan angle only reaches 2.8° at 40 $V_{pp}$. The hysteresis also shifts the main resonance frequency at this point to 141.73 kHz and 145.95 kHz for the down-sweep and up-sweep, respectively. Six scanning micromirrors of identical design are tested from a single fabricated batch and show a standard deviation of approximately 7% of the resonance frequencies for each of the three scanning modes that are evaluated (see Fig. 2C). The frequency deviation for a given mirror appears to be consistent across all modes, which indicate size tolerances through the fabrication process. When comparing this variation to the resonance width of the investigated modes, having a FWHM of approximately 0.3% of the resonance frequency, this is problematic for synchronization purpose when using multiple scanners as it can be difficult to find pairs of scanners with compatible resonance frequencies, i.e. resonance frequencies close enough that both scanners can be actuated at the same frequency, or at frequencies that are desirable for Lissajous addressing in 2 dimensions. The same issue would be present when using the superposition of two orthogonal resonance modes of a single scanner for Lissajous addressing, with tight constraints on the required frequency spacing between both modes.

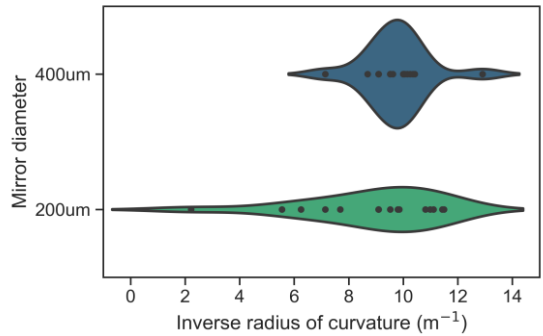

*Figure 3: Radius of curvature distribution of fabricated MEMS scanner in one fabrication run.*

The surface profile of the scanners is characterised using a white-light interferometer (Wyko NT 1100), using a magnification setting to cover the full mirror diameter and measuring the height profile of each mirror with nanometre precision. Most scanners exhibit a mainly spherical surface profile, with an average curvature of roughly 10 dioptres (see Fig. 3). In this case scanners with 200 µm mirror diameter as well as a scaled-up version with 400 µm mirror diameter are compared. Large differences in curvatures can be observed, especially for the 200 µm scanner, with the smaller plate being more prone to deformation from fabrication process tolerances.

## Tuning actuator characterisation

To evaluate the two types of tuning actuators, each is experimentally characterised based on their temperature profile with an employed DC actuation voltage, as well as their in-plane displacement resulting from Joule heating and thermal expansion.

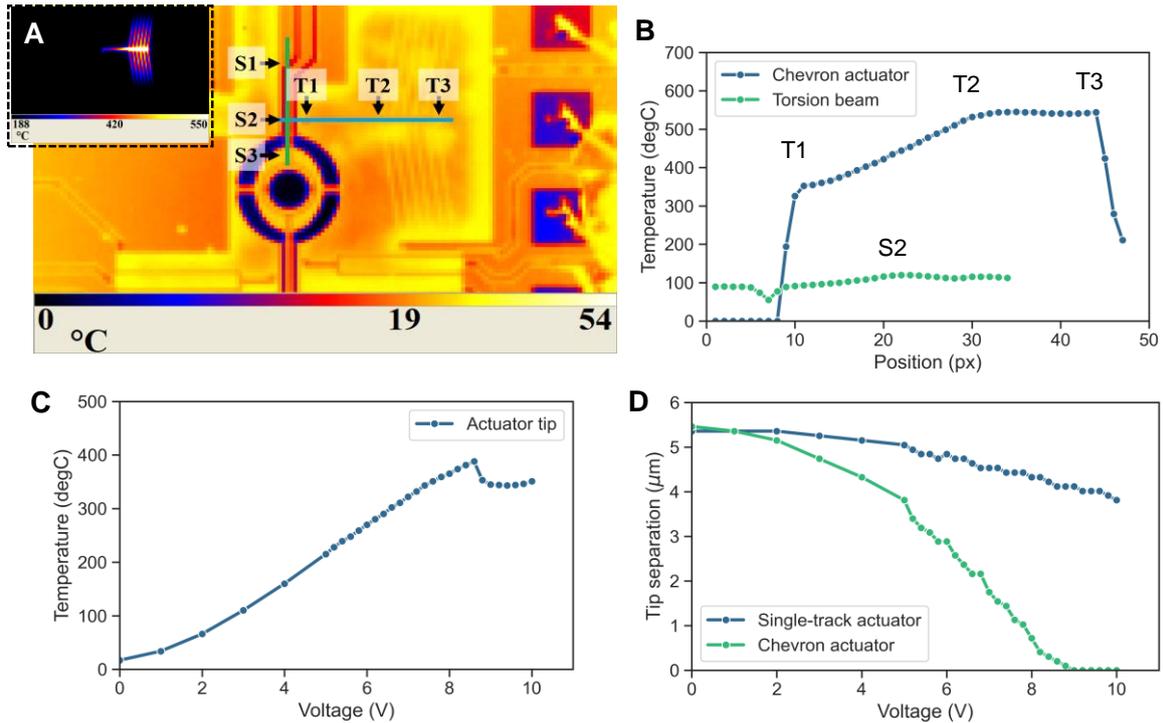

*Figure 4: Measurement of the thermal distribution and response of the chevron frequency tuning actuator. (A) Overview thermal image with 0V actuation to show the profiles over which the temperature is measured on the actuator and scanner torsion beam, and thermal image with actuation of 8V (inset). (B) Temperature distribution for 8V actuation of the chevron tuning actuator, with points highlighted in A marked for reference. (C) Temperature of the actuator tip with increasing tuning voltage and (D) Lateral tip displacement of the chevron actuator with varying DC tuning voltage and comparison with the displacement of the single-track tuning actuators.*

The heating effect produced by the tuning tip is measured using a thermal camera (FLIR SC7000). The surface temperature of the chip can be inferred from the radiation intensity captured by the camera, with the detector calibrated for the emissivity of silicon. Fig. 4A shows an example of a thermal image obtained with the camera without any actuation voltage applied to the chevron actuator. The image highlights the positions of two measurement profiles taken to obtain the temperature of the tuning actuator as well as the main torsion beam of the scanning mirror. At higher actuation voltages the dynamic range of the camera does not allow simultaneous display of both areas due to their temperature difference. The inset in Fig. 4A shows the temperature profile of the chevron actuator at a voltage of 8V, highlighting a maximum temperature reaching around 550°C at the back-end of the actuator. Fig. 4B shows the profile of the two highlighted positions in Fig 4A at the actuation voltage of 8V. Measurements on the chevron tip (see Fig. 4C) indicate that thermal actuation leads to a sharp increase in temperature along the actuation beam, with a temperature maximum on the scanner opposing side of the tuning actuator. Thermal conduction and radiation is translating into a lower increase of the scanner temperature. The maximum temperature of the scanner is reached at the contact point or closest point with the actuator tip, and remains significantly higher than the die temperature at the mirror plate suspension beam. Measurements away from the tuning tip show a significant increase in the temperature of the whole chip, reaching values as high as 50°C. Excessive thermal actuator voltage can lead to melting and destruction of the tuning tips. This typically happens at approximately 12.5 $V_{dc}$. The voltage in any subsequent measurements is therefore kept under 10 $V_{dc}$ as precaution.

The chevron actuator produces enough displacement to make contact with the main suspension beam at approximately 8.6V. In that case, the frequency shift of the scanner resonances is a result of the temperature increase combined with the arguably low mechanical stiffness added by the tip contact. For comparison, the single-track actuators do not possess enough lateral displacement to close the gap to the scanner torsion beam at the maximum tuning voltage of 10V. The lateral tip displacement stays within 1 µm, highlighting that any induced frequency tuning is through thermal effects without addition of mechanical stiffness due to the actuator position.

## Frequency tuning results

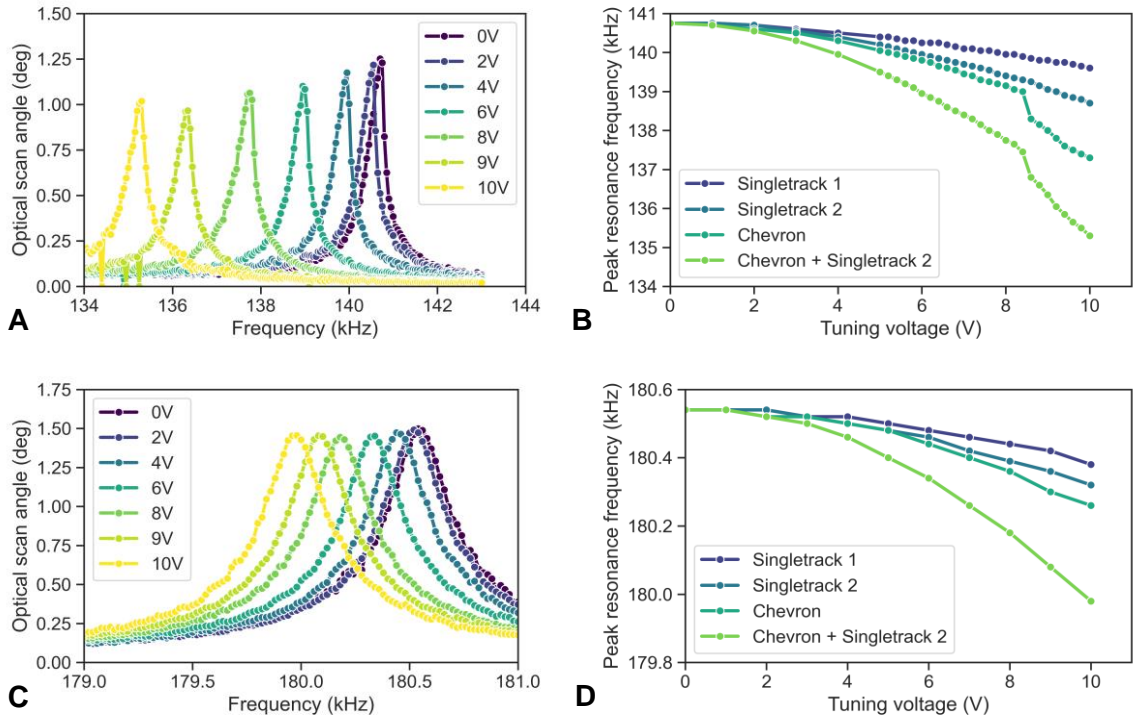

*Figure 5: Frequency tuning measurements using the two type of thermal actuators. (A) Resonance shift for the first orthogonal tip movement mode with scanner actuation with an offset sinusoidal signal with 5Vpp. (B) Comparison of the peak resonance shift for the tip movement when using the three thermal tuning actuators separately or in combination. (C) Resonance shift for the first main tilt movement mode with scanner actuation with an offset sinusoidal signal with 5Vpp. (D) Comparison of the peak resonance shift for the tilt movement when using the three thermal tuning actuators separately or in combination.*

The change of the two main tip and tilt resonance mode frequency responses with active tuning through the two types of tuning actuators is evaluated using again the microscope coupled single point Doppler vibrometer. Fig. 5A shows the evolution of the scan angle frequency response with an applied DC tuning voltage to the chevron and single-track actuator 2 tuning actuators for the first tipping mode of the scanner. The piezoelectric actuator for generating the resonance motion is driven using a 5 Vpp sinusoidal signal with 2.5V offset. With tuning of the peak frequency, only minor changes to the movement Q-factor are visible (from 280 to 220) and the overall resonance scan angle drops by 18% from 1.25° to 1.02°. The evolution of the peak resonance frequency of the first tip mode when using the different tuning actuators separately and in combination is shown in Fig. 5B. When using only a single actuator a maximum frequency shift of 1.15 kHz, 2.05 kHz and 3.45 kHz, can be seen for a tuning voltage of 10 $V_{dc}$ on the single-track 1, single-track 2 and chevron actuator respectively. When combining the actuators this increases to 5.45 kHz for a combination of the chevron actuator and single-track 2 actuator. This directly shows a stronger tuning interaction for the tipping resonance mode for actuators placed closer to the mirror frame. Additionally, for the chevron actuator and its combined actuation with single-track actuator 2 a distinct additional shift in frequency is visible at around 8.6V tuning actuation, which is originating from the contact made between the tuning tip and scanner main torsion beam at that point. Fig. 5C and D show the same frequency response evolution and peak resonance frequency shifts for the first tilt mode of the scanner, again with actuation of a single piezoelectric actuator with a 5 Vpp sinusoidal signal with 2.5 V offset. The resulting tuning range is significantly smaller, with a maximum shift of 0.16 kHz, 0.22 kHz and 0.28 kHz for a tuning voltage of 10 $V_{dc}$ on the single-track 1, single-track 2 and chevron actuator respectively. For a combined tuning of the chevron actuator and single-track actuator 2 the shift increases to 0.56 kHz with 10 $V_{dc}$ tuning. The Q-factor for the tilt resonance has again only a minor change for the maximum tuning range (from 420 to 400) and the overall resonance scan angle drops only marginally by 3% from 1.5° to 1.45°.

## Discussion

Initial characterisation of the fabricated scanners shows variation in mirror profile and frequency response among devices with identical designs. These discrepancies are likely due to fabrication tolerances during the fabrication process, where slight changes in layer deposition and etching can result in unpredicted residual stress within the scanner structure and mirror plate deformation. This itself has in the case of the investigated scanners resulted in a mode frequency variation of ~7%.

Tuning of the resonance frequencies of the two orthogonal tip and tilt modes has been achieved with a maximum shift of 5.45 kHz (3% of the resonance frequency). Generally, tipping modes are found to be significantly more affected by the investigated on-chip electrothermal tuning than tilting modes. This is likely due to the tuning actuation having a localised heat effect, with the main temperature change present at the main torsion beam of the scanner, near to where the tuning tips are positioned. The main torsion beam is more involved in the motion of tipping modes, as can be seen from the mode shapes in Fig. 2A; it undergoes an out-of-plane flexion, particularly in the first tipping mode, whereas it is nearly immobile in tilting modes.

Fig. 5B and D both show that the frequency shift resulting from thermal effects of the tuning actuators evolve with the square of the tuning voltage; this suggests an increase in peak temperature, and thus a decrease in beam stiffness, proportional to the electrical power dissipated through Joule heating, which is indeed proportional to the square of the applied voltage. Tuning through thermal contact was also observed, manifested by an additional sharp reduction of the resonance frequency at the point contact is made for the tip movement shown in Fig. 5B. However, the frequency diminution goes against the expected frequency increase resulting from a stiffer beam by adding a clamping force to the suspension. Instead, it is possible that contact facilitates heat transfer to the scanner as indicated by the temperature drop of the actuator tip in Fig. 4C above 8.6 V, raising its temperature further and overcompensating any mechanical stiffness increase. Tip contact does not affect the scanning amplitude significantly compared to the non-contact heat induced shifts, suggesting that the damping and effective mass of the resonating MEMS remains mostly unchanged.

Having shown that the presented on-chip thermal tuning has a significantly different influence on the two main orthogonal scanning resonance modes of the device leads to an interesting potential compared to mass tuning of the mirror surface, which influences all movement modes. When driving the scanner simultaneously with multiplexed frequencies corresponding to the two tip/tilt resonances to create a 2D Lissajous scan pattern, it becomes possible to easily tune the frequency ratio between both modes. This ability will allow control over the pattern repetition frequency and time to address a full 2D scene, as well as the density of scan pattern [13], [14]. Using the frequency tuning approach and the frequencies of the scanner presented here allows for a maximum pattern repetition frequency of 20 kHz, while the highest fill factors having 173 x 132 cross-over points of the Lissajous pattern are possible with a pattern repetition frequency of 1 kHz, all the while keeping the maximum available scan angles.

## Conclusion

A fast piezoelectric scanning micromirror with tip/tilt resonance frequencies above 130 kHz and on-chip frequency tuning through electrothermal actuators has been developed and characterized. The two fundamental tip/tilt resonance movement modes of the 200 µm diameter mirror are at 141 kHz and 180 kHz, with optical scan angles of 10° and 7° with 40V actuation. Using two variations of electrothermal actuators fabricated adjacent to the scanner suspension whose heated tuning tips are positioned within 10 µm of the main torsion beam of the scanner allows for frequency tuning of one of the modes by up to 5.45 kHz while the second mode changes by only 500 Hz. Thermal conduction and radiation from the tip of the tuning actuators create global and local heating of the main torsion beam, with spring softening of the beams leading to a reduction of resonance frequencies of modes which are dominated by movement of the main torsion beams, while movement modes without impact by the torsion beam show only minimal tuning. The combined actuation of both modes at the same time will allow high frequency Lissajous 2D scanning from a single device, with the mode selective tuning allowing the selection of pattern repetition frequencies and fill factors depending on the frequency ratio between the modes, as well as compensation of frequency shifts due to fabrication tolerances.

## Acknowledgements

We want to thank Paul Griffin and Erling Riis for fruitful discussions on MEMS requirements and application directions. We acknowledge funding for an NPL studentship, and funding from the UK Engineering and Physical Sciences Research Council (grant EP/S032606/1) and UK Royal Academy of Engineering (Engineering for Development Fellowship scheme RF1516/15/8). For the purpose of open access, the authors have applied a Creative Commons Attribution (CC BY) licence to any Author Accepted Manuscript version arising from this submission.

## Data availability

All research data and materials supporting this publication can be accessed at DOI: https://doi.org/10.15129/c589d644-e614-4c0c-80b6-5af7fe786492

## Competing interests

The authors declare no competing interests